\documentclass[
superscriptaddress,
twocolumn,
amsmath,amssymb,
aps,
pra,
]{revtex4-2}

\usepackage{amsmath,amssymb,amsfonts,braket,bm}
\usepackage{graphicx}
\usepackage{enumerate}

\usepackage{qcircuit}
\usepackage{hyperref}

\begin{document}
\preprint{}

\title{A Quantum-Resistant Photonic Hash Function}

\author{Tomoya Hatanaka}
\affiliation{Blocq, Inc., Toranomon Hills Business Tower 1-17-1 Toranomon, Minato, Tokyo 105-6416 Japan}

\author{Rikuto Fushio}
\affiliation{Blocq, Inc., Toranomon Hills Business Tower 1-17-1 Toranomon, Minato, Tokyo 105-6416 Japan}

\author{Masataka Watanabe}
\affiliation{Graduate School of Informatics, Nagoya University, Nagoya 464-8601, Japan}
\affiliation{Blocq, Inc., Toranomon Hills Business Tower 1-17-1 Toranomon, Minato, Tokyo 105-6416 Japan}

\author{William J. Munro}
\affiliation{Okinawa Institute of Science and Technology Graduate University, Okinawa 904-0495, Japan}

\author{Tatsuhiko N. Ikeda}
\email{tatsuhiko.ikeda@riken.jp}
\affiliation{RIKEN Center for Quantum Computing, Wako, Saitama 351-0198, Japan}

\author{Sho Sugiura}
\email{sho.sugiura@blocqinc.com}
\affiliation{Blocq, Inc., Toranomon Hills Business Tower 1-17-1 Toranomon, Minato, Tokyo 105-6416 Japan}

\begin{abstract}

We propose a quantum hash function based on Gaussian boson sampling on a photonic quantum computer, aiming to provide quantum-resistant security. Extensive simulations demonstrate that this hash function exhibits strong properties of preimage, second preimage, and collision resistance, which are essential for cryptographic applications. Notably, the estimated number of attempts required for a successful collision attack increases exponentially with the mode counts of the photonic quantum computer, suggesting robust resistance against birthday attacks. We also analyze the sampling cost for physical implementation and discuss potential applications to blockchain technologies, where the inherent quantum nature of the hash computation could provide quantum-resistant security. The high dimensionality of the quantum state space involved in the hashing process poses significant challenges for quantum attacks, indicating a path towards quantum security. Our work lays the foundation for a new paradigm of quantum-resistant hashing with applications in emerging quantum-era information systems. 
\end{abstract}

\maketitle

\section{Introduction}
\label{sec:introduction}

Quantum computers can solve complex problems, but their advent may inadvertently spawn new quandaries that society must confront~\cite{Bacon2010}.
Although quantum computers promise to solve specific problems more efficiently and with reduced energy consumption, their potential to provide malicious actors with a powerful new tool for breaking cryptographic security poses a significant concern~\cite{Shor2006}.
This will be a particular problem for cryptocurrency (or, more generally, blockchains), where encrypted information can be accessed publicly -- the hackers can collect information beforehand and decrypt it once quantum computers are available~\cite{Aggarwal2018}.
As a consequence, it is desirable to update some of the current blockchains to have quantum resistance as soon as possible~\cite{Fernandez-Carames2020}.
It would be an immediate and simple strategy to increase the number of output bits in existing hash functions, which are then believed to be quantum-resistant.
However, the quantum resistance of existing hash functions hinges on the sheer fact that no one yet knows the quantum algorithm that solves them backward.
Moreover, it is also a possibility that the increase in the number of bits results in the increase in hash computation time and energy usage.
Because of these, as the security requirement rises along with the sophistication of quantum computing technology, there is a possibility that such a simple strategy no longer works in the near future. 
It would therefore be desirable to create a new hash function to tackle this problem.

There have been general theoretical proposals for designing a hash function using quantum computers~\cite{Kashefi2005,Ablayev2013,Ablayev2018}. 
To give a concrete example, a hash function was proposed based on quantum walks on a universal digital quantum computer~\cite{Banerjee2024, Hou2023, Li2018, Yang2016, Zhou2022}, while the necessary quantum resources for its physical implementation remain yet to be fully understood. 
Meanwhile, photonic quantum computers also offer another promising candidate platform~\cite{Arrazola2021, Eli-Bourassa2021, O-Brien2009}.
Indeed a quantum hash function was proposed based on coarse-grained boson sampling~\cite{Nikolopoulos2016,Nikolopoulos2019}, which is realized on a photonic quantum computer, or more specifically a linear optical interferometer with photon detectors~\cite{Calkins2013, Hadfield2009, Lita2008}.
Such a task is believed to achieve quantum advantage~\cite{Aaronson2011,Brod2019} and as such can be considered truly quantum. However, these hash functions are still challenging in terms of experimental realization, as they use states with fixed photon numbers as inputs of boson sampling.

In this paper, we propose a new quantum hash function using Gaussian boson sampling (GBS), which is a type of boson sampling whose inputs are Gaussian states instead of Fock states.
The construction is so that it makes use of the quantum advantage as well as experimentally realizable using current technology.
Concretely, we first translate the input message into a Gaussian state, which then go through a linear optical interferometer.
Then the three-body correlation functions of the output state are measured, after which they are translated into the output string.
Because we input Gaussian states into the linear interferometer, the experimental realization of GBS is less challenging; Indeed it has already been accomplished in~\cite{Zhong2020,Zhong2021,Madsen2022,Dellios2023,Deng2023}.
It is also believed to retain quantum advantage like its Fock-state counterpart~\cite{Hamilton2017,Kruse2019}. 
Therefore, the use of GBS makes our hash function hard to simulate classically yet experimentally realizable with current technology.
This contrasts with~\cite{Pai2023}, which proposed a hash function that determines output bits based on the classical laser light power in each mode of the linear interferometer without utilizing GBS. As a result, it does not achieve a quantum advantage and is essentially equivalent to classical light-based computing~\cite{Singh2024}.

We then systematically evaluate the security properties of our quantum hash function.
In particular, through extensive simulations, we conclude that it demonstrates preimage, second preimage, and collision resistance.
Importantly, the estimated number of attempts required for a successful collision attack is found to increase exponentially with the number of modes, indicating robust collision resistance. 
In addition to security analysis, we examine the sampling cost associated with physically implementing the quantum hash function. 
The required number of measurement shots to achieve a target precision is largely independent of the number of modes. 
This provides practical insights into the quantum resources needed for real-world deployment. 
Also, the involvement of quantum measurements in the hashing process is expected to provide inherent resistance against quantum attacks.

Our paper is organized as follows. Section \ref{sec:photonichash} describes the proposed quantum hash function based on GBS. Section \ref{sec:simulations} presents detailed security analyses through benchmark simulations and examines the sampling cost for physical implementation. In Section IV, we discuss how our quantum hash function makes blockchains more secure as potential applications. Finally, Section V summarizes our study and discusses future outlooks.

\section{Photonic hash function}
\label{sec:photonichash}
\begin{figure}
    \includegraphics[width=\columnwidth]{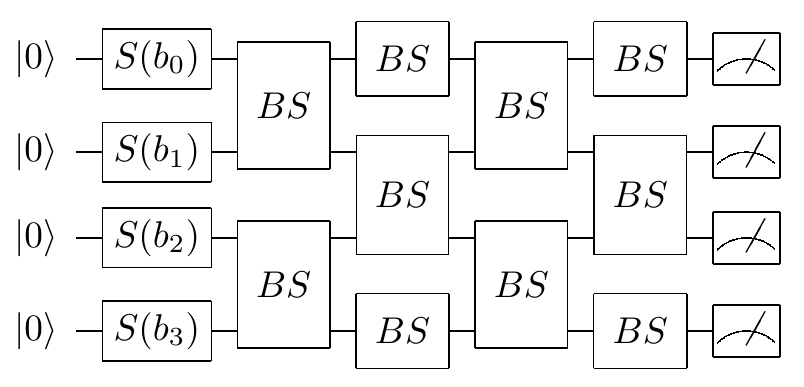}
  \caption{Illustration of the interferometer for $N=4$ and $d=4$. The periodic boundary conditions are imposed on the boundary modes.}
  \label{fig:quantum_circuit}
\end{figure}

Our photonic hash function can be defined on boson samplers and their variants.
Among such variations, we focus on the GBS, as the input states are easier to prepare with current technology. 
We consider an $N$-mode linear interferometer to implement an $N$-bit hash function.
The configuration of the interferometer is constructed with beamsplitters arrayed in a brickwork structure as in Fig.~\ref{fig:quantum_circuit}.
We also assume the periodic boundary condition on the both ends of the modes but other similar structures or boundary conditions almost equally work as discussed in Appendix~\ref{app:rave}.
It encodes a product of unitary matrices, 
\begin{align}\label{eq:mU}
    \mathcal{U}=\prod_{l=0}^{d-1}\mathcal{U}_l,
\end{align}
where $d$ is the depth of the interferometer, and $\mathcal{U}_l$ is one layer of bricks given by the action of beamsplitters connecting neighboring modes, alternating their connections in terms of the parity of $l$:
\begin{align}
    &\mathcal{U}_l=
    \begin{cases}
        \displaystyle\prod_{j=0}^{N/2-1}\mathrm{BS}_{2j,2j+1}(\theta_j^{(l)},\phi_j^{(l)}) & (\text{$l$: even})\\
        \displaystyle\prod_{j=0}^{N/2-1}\mathrm{BS}_{2j+1,2j+2}(\theta_j^{(l)},\phi_j^{(l)})& (\text{$l$: odd}),
    \end{cases}
\end{align}
\begin{align}
    \mathrm{BS}_{j,j'}(\theta,\phi)\equiv\exp\left[ \theta(e^{i\phi}a_{j}a_{j'}^\dag - e^{-i\phi}a_{j}^\dag a_{j'})\right].
\end{align}
Here $a_j$ ($a_j^\dag$) represents the photon annihilation (creation) operators for the $j$-th mode, where the index $j=N$ should be understood as $j=0$, due to the periodic boundary condition we impose.

As we want the resulting unitary to be random, $\phi_j^{(l)}$ will be chosen independently from the uniform distribution over the interval $0\le \phi<2\pi$. 
On the other hand, $\theta_j^{(l)}$ are chosen independently from the normal distribution with mean $\pi/4$ and standard deviation $\pi/16$.
This is to enhance mode mixing, noting that $\theta=\pi/4$ gives a 50-50 beamsplitter. The $\pi/4$ periodicity in $\theta$ allows us to identify a negative $\theta$ as a positive one using the modulo structure.
Like other random interferometers, the constructed interferometer is so complex that the linear relation between the input and output photon modes are close to the Haar random unitary matrices (see Appendix~\ref{app:rave} for justification).
In blockchain applications, these parameters are chosen at random initially and then will be publicly shared for all parties.

Let us now explain how to use the interferrmoeter in order to compute the hash output.
First of all, we prepare an input binary bitstring (message) $\bm{b}=b_{N-1}b_{N-2}\dots b_0$, where $b_j$ is either $0$ or $1$.
There are $2^N$ distinct input bitstrings, which we write as $\bm{b}^{(i)}$ ($i=0,1,\dots,2^N-1$). We let $\mathcal{B}$ denote their set $\mathcal{B}=\{\bm{b}^{(i)}\}_{i=0}^{2^N-1}$.
One of the inputs $\bm{b}$ is embedded into an array of squeezed vacuum,
\begin{align}\label{eq:instate}
    &\ket{\Psi_\mathrm{in}(\bm{b})} = \prod_{j=0}^{N-1} S_j(b_j) \ket{0},
\end{align}
where
\begin{align}\label{eq:squeezingop}
    &S_j(r)\equiv \exp\left[\frac{r}{2}( a_j^2- (a_j^\dag)^2)\right]
\end{align}
denotes the squeezing operator for the $j$-th mode, with $r$ being a real squeezing parameter.
Equation~\eqref{eq:instate} means that the squeezing parameter is $r=b_j=0$ or $1$ at each mode $j$, depending on the input $\bm{b}$. Here, $r=0$ squeezing means no operation, whereas $r=1$ squeezing is a typical experimental value in GBS experiments (e.g., $r\approx 1.1$ in~\cite{Madsen2022}).

The input state is then fed into the interferometer
\begin{align}
    \ket{\Psi_\mathrm{out}(\bm{b})} = \mathcal{U}\ket{\Psi_\mathrm{in}(\bm{b})}.
\end{align}
We conduct photon number measurements on the state $\ket{\Psi_\mathrm{out}(\bm{b})}$ to get the hash output.
Concretely, we first measure the nearest-neighbor three-mode correlation functions $\mu_j$, defined as 
\begin{align}
    \mu_j(\bm{b}) &\equiv \langle  C_{j} \rangle_{\bm{b}}, \\
    C_{j} &\equiv n_j n_{j+1}n_{j+2}, \label{eq:3body}
\end{align}
where $n_j \equiv a_j^\dag a_j$ is the photon number operator of the $j$-th mode. 
It is important to mention here that we consider three-mode correlations rather than one- or two-mode correlations as those two can be classically computed relatively easily~\cite{Dodonov1994}.
Although we focus on the three-mode correlations, the following discussions equally apply to other $m$-mode correlations, and $m=O(N)$ would make the classical computations very difficult~\cite{Cardin2023}.

We introduce a decimal parameter $k$ in converting $\{\mu_j(\bm{b})\}_{j=0}^{N-1}$ into an output bitstring (i.e., hash value) $\bm{c}=c_{N-1}c_{N-2}\dots c_0$ with $c_j$ being either 0 or 1.
For a given $k$, we convert each of $\mu_j$ into a binary $c_j$ as follows\footnote{A similar idea was also used in quantum walk hash functions~\cite{Hou2023, Li2018, Yang2016, Zhou2022}, and 10 can be replaced by other integers.},
\begin{align}\label{eq:mu_to_c}
    c_j = \begin{cases}
        0 & (\text{$k$-th decimal of $\mu_j(\bm{b})$ is even}),\\
        1 & (\text{$k$-th decimal of $\mu_j(\bm{b})$ is odd}),
    \end{cases}
\end{align}
which is also expressed as $c_j \equiv \left\lfloor 10^k \mu_j\right\rfloor \mod 2$ using the floor function $\lfloor \cdots\rfloor$.
Note that for two special inputs $\bm{b}=00\dots0$ and $11\dots1$, we tirivially have $\mu_j=0$ for all $j$, and hence $\bm{c}=00\dots0$. 
These two inputs are to be removed in practical applications.

A trade-off relation exists between smaller and larger $k$.
Roughly speaking, a larger $k$ makes the hash value expectedly more sensitive to the input because a smaller fraction of $\mu_j$ is used, and this is favorable for hash functions.
However, a larger $k$ also means that $\mu_j$ must be obtained precisely up to $k$-th decimal for the hash function to be deterministic.
This means that both systematic and statistical errors of the measurement outcomes $\mu_j$ must be below $10^{-k}$, which becomes more demanding for larger $k$. In the following section, we will discuss how hash function properties depend on $k$ with benchmark simulations.

Combining the above procedure, we obtain the hash function in the form of
\begin{align}
    \bm{c} = \mathrm{Hash}_{\bm{\theta},\bm{\phi},k}(\bm{b}).
\end{align}
Here $\bm{\theta}$ and $\bm{\phi}$ characterize the interferometer, and $k$ determines which decimal of $\mu_j$ is adopted in the hash value. Below we will use a notation $\mathcal{C}$ to denote the set of the hash values: $\mathcal{C}=\{\bm{c}^{(i)}=\mathrm{Hash}_{\bm{\theta},\bm{\phi},k}(\bm{b}^{(i)})\}_{i=0}^{2^N-1}$.

In computing the hash function using a physical interferometer, one must estimate the quantum expectation values in $\mu_j$ based on sampling. Each measurement outcome in the photon detectors is a set of photon numbers at each mode $j$. After taking $N_\mathrm{shot}$ measurement results, we estimate expectation values in \eqref{eq:3body} as sample averages, and hence there is estimation errors that scale as $\propto N_\mathrm{shot}^{-1/2}$, as long as systematic errors are neglected. For the hash function to be deterministic and not to be affected by the estimation errors, $N_\mathrm{shot}$ has to be such that $N_\mathrm{shot}^{-1/2} \lesssim 10^{-k}$. In the following section, we estimate the required $N_\mathrm{shot}$ with benchmarking simulations.

\section{Benchmark simulations}
\label{sec:simulations}
Now let us explore the essential characteristics and performance metrics of the quantum hash function, which are critical in evaluating their suitability for cryptographic applications. A quantum hash function, unlike its classical counterpart, leverages the principles of quantum mechanics to achieve enhanced security features and computational efficiency. To systematically assess these functions, we focus on three fundamental aspects: unpredictability, choice of decimal parameter $k$ and sampling cost.
We assume that the quantum expectation values $\mu_j$ are obtained without statistical errors in Sections~\ref{sec:unpredictability} and \ref{sec:select_decimal}, and discuss the errors in Section~\ref{sec:samplingcost}.

Throughout this section, we set the depth of the quantum circuit equal to the number of modes, $d=N$, and fix a typical set of the random numbers $\theta,\phi$ for the beamsplitters. For brevity, we thus abbreviate $\mathrm{Hash}_{\bm{\theta},\bm{\phi},k}\equiv\mathrm{Hash}_{k}$ in the following discussions.

\subsection{Unpredictability of Quantum Hash Function}\label{sec:unpredictability}
In this subsection, we discuss three essential properties that a secure hash function must satisfy: preimage resistance, second preimage resistance, and collision resistance~\cite{Rogaway2004}. These properties ensure that the hash function is robust and secure in cryptographic applications~\cite{Al-Kuwari2011}.

\begin{itemize}
    \item \textit{Preimage Resistance}: Given a hash output $\bm{c}$, it should be computationally difficult to find any input $\bm{b}$ such that $\mathrm{Hash}(\bm{b}) = \bm{c}$. This ensures that an attacker cannot reverse the hash to obtain the original input.
    
    \item \textit{Second Preimage Resistance}: Given an input $\bm{b}$, it should be difficult to find another input $\bm{b}^\prime \neq \bm{b}$ such that $\mathrm{Hash}(\bm{b}) = \mathrm{Hash}(\bm{b}^\prime)$. This prevents attackers from finding alternative inputs that produce the same hash value as a given input.
    
    \item \textit{Collision Resistance}: It should be difficult to find two different inputs $\bm{b} \neq \bm{b}^\prime$ that produce the same hash output, i.e., $\mathrm{Hash}(\bm{b}) = \mathrm{Hash}(\bm{b}^\prime)$. This ensures that the hash function is resilient to attacks that attempt to find input pairs with matching hashes.
\end{itemize}

Direct measurement of these properties is often challenging. Therefore, we measure \textit{confusion} as a proxy for preimage resistance and \textit{diffusion} as a proxy for second preimage resistance~\cite{Hou2023, Li2018, Yang2016, Zhou2022}. Additionally, we estimate collision resistance by assessing the resilience of the hash function against a \textit{birthday attack}, which is a method to find hash collisions~\cite{Wiener2005}.

\subsubsection{Confusion}
Confusion in a hash function refers to the property that creates a complex and nonlinear relationship between the input and output~\cite{Hou2023, Li2018, Yang2016, Zhou2022}. We assess this property by (i) the degree of confusion and (ii) the uniformity of confusion. The former measures how significantly the output of the quantum hash function changes in response to various inputs, whereas the latter examines whether the bit changes in the output are uniformly distributed.

First, we evaluate the degree of confusion.
Let us begin by quantifying correlations between an input message $\bm{b}^{(i)}=b_{N-1}^{(i)}b_{N-2}^{(i)}\dots b_0^{(i)}$ $(i=0,1,\dots,2^N-1)$ and its output $\bm{c}^{(i)}=c_{N-1}^{(i)}c_{N-2}^{(i)}\dots c_0^{(i)} = \mathrm{Hash}_{k}(\bm{b}^{(i)})$.
For this purpose, we consider the Hamming distance
\begin{align}
    d_k(\bm{b},\bm{c})
    &=
    \sum_{j=0}^{N-1} (b_j \oplus c_j),
\end{align}
where the operator $\oplus$ denotes the exclusive OR (XOR) operation, i.e. $b\oplus c=1$ if $b\neq c$ and $b\oplus c=0$ if $b=c$. 
We then average the Hamming distances over all the $2^N$ inputs and divide it by $N/2$ for normalization,
\begin{align}\label{eq:hamming_distance}
    D_k(\mathcal{B}, \mathcal{C}) 
    &=
    \frac{2}{N} \left(\frac{1}{2^N} \sum_{i=0}^{2^N-1}d_k(\bm{b}^{(i)},\bm{c}^{(i)})\right),
\end{align}
The normalization factor $2/N$ makes $D_k(\mathcal{B},\mathcal{C})$ ranges from 0 to 2, and these two limiting values imply a kind of trivial correlations between the inputs and outputs: $\bm{c}^{(i)}=\bm{b}^{(i)}$ for every $i$ if $D_k=0$, and $\bm{c}^{(i)}=\neg\bm{b}^{(i)}$ for every $i$ if $D_k=2$ ($\neg$ denotes flipping a bit). In between ($D_k=1$), the inputs and outputs are interpreted to be ideally uncorrelated.

\begin{figure*}
\centering
        \includegraphics[width=\columnwidth]{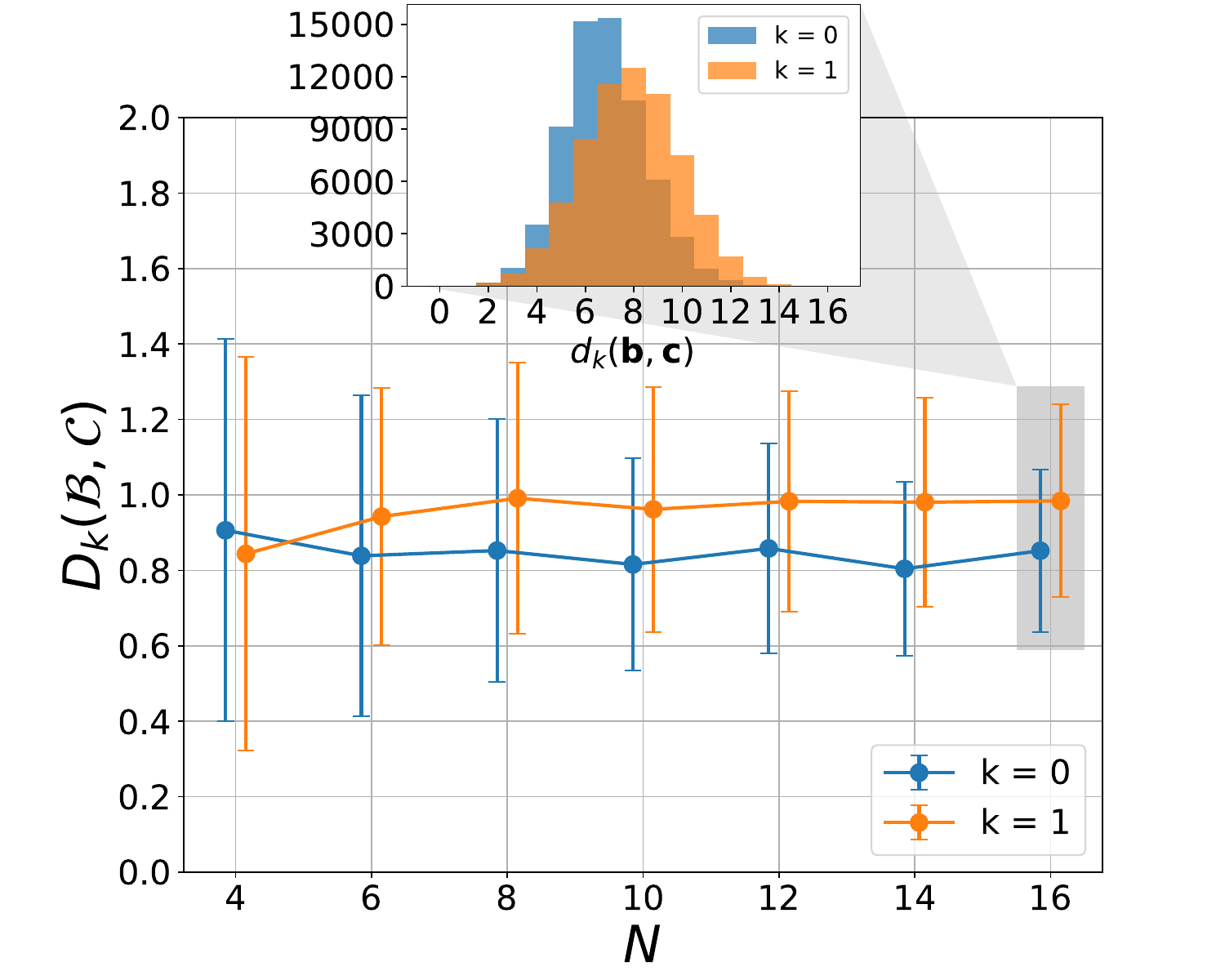}
        \includegraphics[width=\columnwidth]{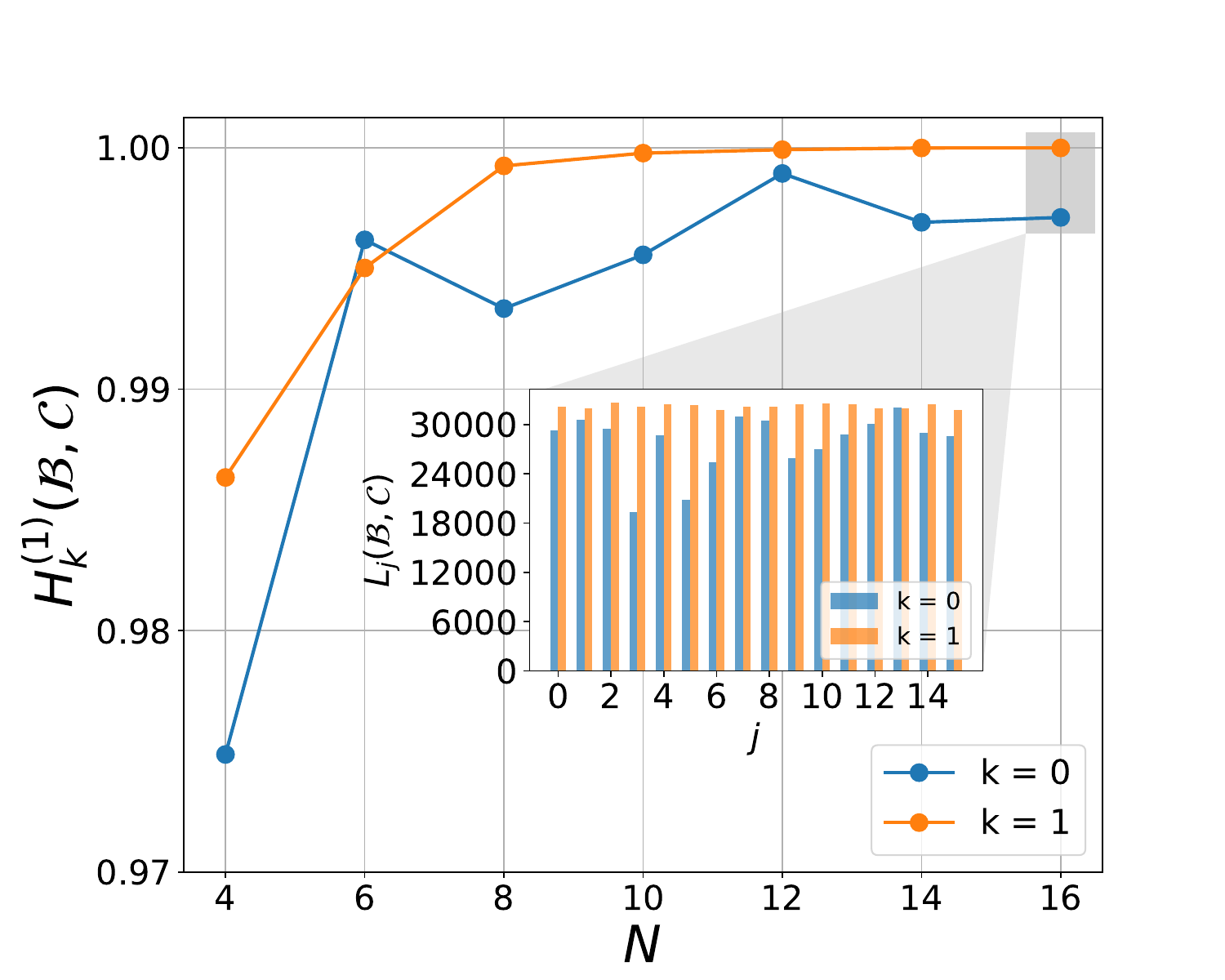}
    \put(-500,190){(a)}
    \put(-250,190){(b)}
    \caption{Confusion analysis of the quantum hash function properties. (a) The degree of confusion shown by the normalized averaged normalized Hamming distance $D_k(\mathcal{B}, \mathcal{C})$ for various $N$. The error bars show the standard deviation of the normalized Hamming distance.
    The inset presents the histogram of $d_k(\bm{b},\bm{c})$ for $N=16$.
    \
    (b) The uniformity of confusion given by the normalized Shannon entropy $H_k^{(1)}(\mathcal{B}, \mathcal{C})$ for various $N$.
    The inset shows a snapshot for $N=16$, illustrating the distribution of the numbers $L_j(\mathcal{B}, \mathcal{C})$ of bit changes at bit position $j$. 
    }
    \label{fig:confusion}
\end{figure*}

Fig.~\ref{fig:confusion}(a) shows the normalized average Hamming distance $D_k(\mathcal{B}, \mathcal{C})$ for various $N$. The error bars in the figure represent the standard deviation normalized by $N/2$, providing a measure of the variability in bit changes as the input is altered. The results indicate that for $k = 1$, the normalized average Hamming distance asymptotically approaches the ideal value as $N$ increases. However, for $k=0$, the distance remains slightly below the ideal value of $1$, reflecting a less consistent degree of bit change. To provide a specific snapshot of the degree of confusion, we analyzed the distribution of bit changes for the case where $N=16$. The histogram in the inset of Fig.~\ref{fig:confusion}(a) shows Hamming distance $d_k(\bm{b},\bm{c})$, revealing that the distribution exhibits a Gaussian shape with a mean ideally centered at $N/2$ (which is $8$ for $N=16$). This indicates that the degree of confusion remains relatively stable regardless of the value of $k$, as the histogram shows little variation in the number of bit changes across different values of $k$.

Second, we examine the uniformity of confusion. 
We begin by counting how many times each $j$-th bit is flipped ($b_j\oplus c_j=1$) or not ($b_j\oplus c_j=0$) when all possible $2^N$ messages are put into our quantum hash function,
\begin{align}
    L_j(\mathcal{B}, \mathcal{C}) &= \sum_{i=0}^{2^N-1} (b_j^{(i)} \oplus c_j^{(i)}).
\end{align}
We ask how uniformly these counts $L_j$ distribute over the modes $j$, or examine the uniformity of the probability distribution
\begin{align}
    P_j(\mathcal{B}, \mathcal{C}) &= \frac{L_j(\mathcal{B}, \mathcal{C})}{\sum_{j^\prime=0}^{N-1} L_{j^\prime}(\mathcal{B}, \mathcal{C}) }.
\end{align}
Next we quantify the uniformity of this distribution by the normalized Shannon entropy
\begin{align}\label{eq:shannon_entropy}
    H_k^{(1)}(\mathcal{B}, \mathcal{C}) &=  \frac{-\sum_{j=0}^{N-1} P_j(\mathcal{B}, \mathcal{C})\log_2 P_j(\mathcal{B}, \mathcal{C})}{\log_2 N},
\end{align}
where the denominator normalizes the entropy so that $H_k^{(1)}(\mathcal{B},\mathcal{C})$ ranges from 0 (maximally nonuniform) to 1 (maximally uniform) and enables us to compare with different $N$.
Here, the notation $H^{(1)}$ specifically indicates that this entropy is the Shannon entropy, which corresponds to the R\'{e}nyi entropy in the limit as the order parameter approaches $1$. This reflects the fact that Shannon entropy is a special case of R\'{e}nyi entropy, providing a measure of uncertainty or randomness in the bit change distribution.

Fig.~\ref{fig:confusion}(b) shows numerically obtained $H_k^{(1)}(\mathcal{B}, \mathcal{C})$ measuring the uniformity of the bit changes across all bit positions for different modes. The results reveal that for values of $k = 1$, the entropy values approach the ideal value of $1$ as $N$ increases, indicating a uniform distribution of bit changes. This suggests that the quantum hash function exhibits a consistent level of uniformity in bit changes across the bit positions as $N$ grows. However, for $k=0$, occasional deviations from the ideal value are observed, suggesting a less uniform distribution of bit changes in some cases. To provide a specific snapshot of the uniformity of confusion, we analyzed the bit changes for the case where $N=16$. The bar chart in the inset of Fig.~\ref{fig:confusion}(b) shows total number of bit changes $L_j(\mathcal{B}, \mathcal{C})$ at bit position $j$ for $N=16$. The chart highlights that the distribution of bit changes is uniformly spread across different bit positions, with each bit position showing changes around $2^{N-1}$. Although some bias is observed for $k=0$, the distribution of bit changes is uniformly spread across different bit positions for $k=1$, with each bit position showing changes around $2^{N-1}$. This consistency across all bit positions for these values of $k$ further supports the entropy analysis and provides an example of the ideal uniform distribution of bit changes, reinforcing the robustness of the quantum hash function under varying conditions.

\subsubsection{Diffusion}
Diffusion (the sensitivity and avalanche effect) in the context of hash functions refers to how significantly the hash value changes against a local change in the input~\cite{Hou2023, Li2018, Yang2016, Zhou2022}. Like we did for confusion, we assess (i) the degree and (ii) the uniformity of confusion, respectively. To test diffusion quantitatively, we conduct the diffusion test method used in~\cite{Yang2016}, which goes in the following steps:
\vspace{3mm}
\begin{enumerate}
    \item Compute the hash value $\bm{c} = \mathrm{Hash}_{k}(\bm{b})$ of an input value $\bm{b} \in \mathcal{B}$
    \item Flip a randomly chosen bit of $\bm{b}$ to obtain the modified input value $\bm{b}^\prime$ and compute the hash value $\bm{c}^\prime = \mathrm{Hash}_{k}(\bm{b}^\prime)$
    \item Compute the Hamming distance between $\bm{c}$ and $\bm{c}'$, i.e. $d_k(\bm{c},\bm{c}')$ and whether a bit change occurred at the $j$-th bit position, i.e. $c_j \oplus c_j^\prime$
    \item Repeat from 1) to 3) for $\forall \bm{b} \in \mathcal{B}$
\end{enumerate}

\begin{figure*}
    \includegraphics[width=\columnwidth]{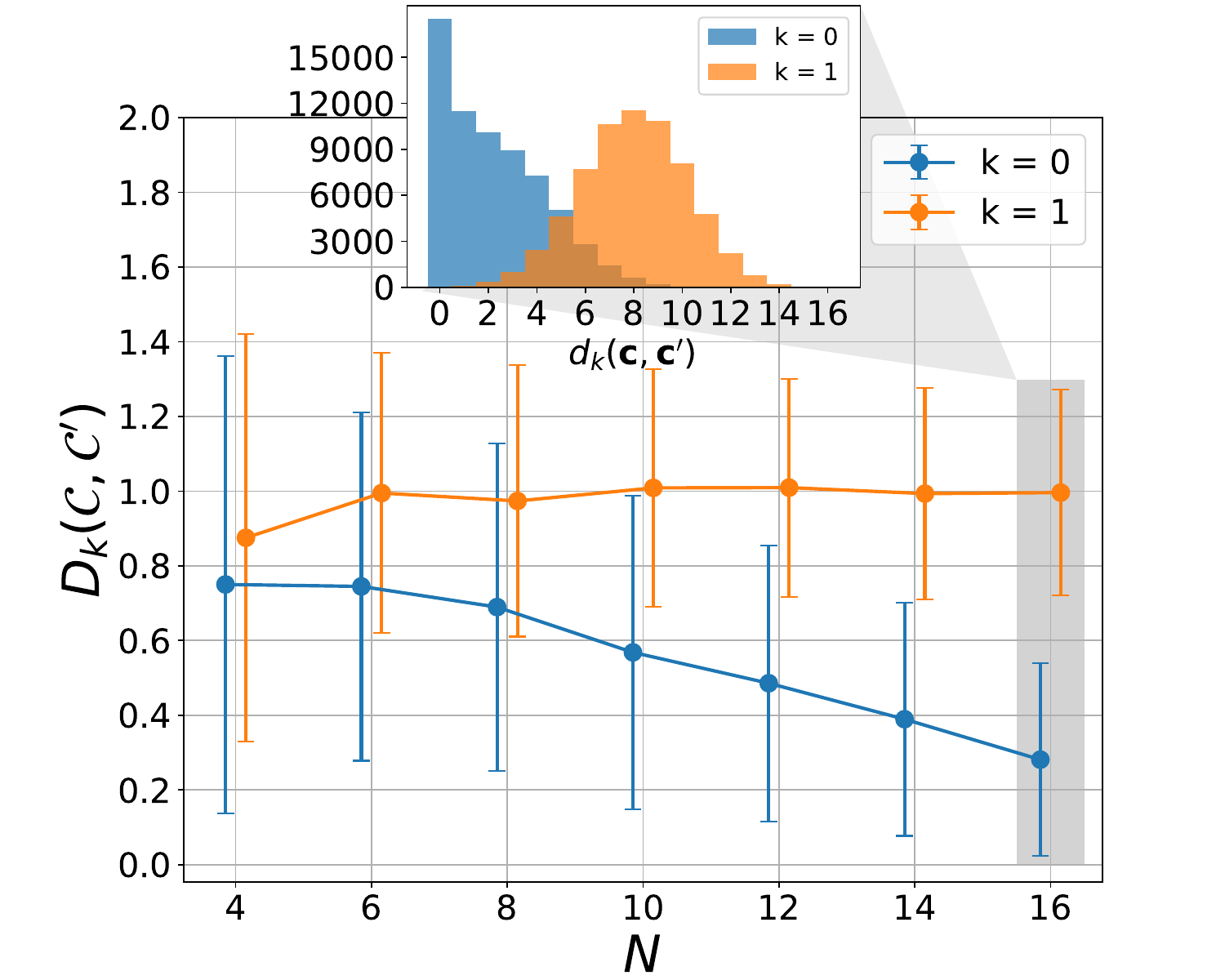}
    \includegraphics[width=\columnwidth]{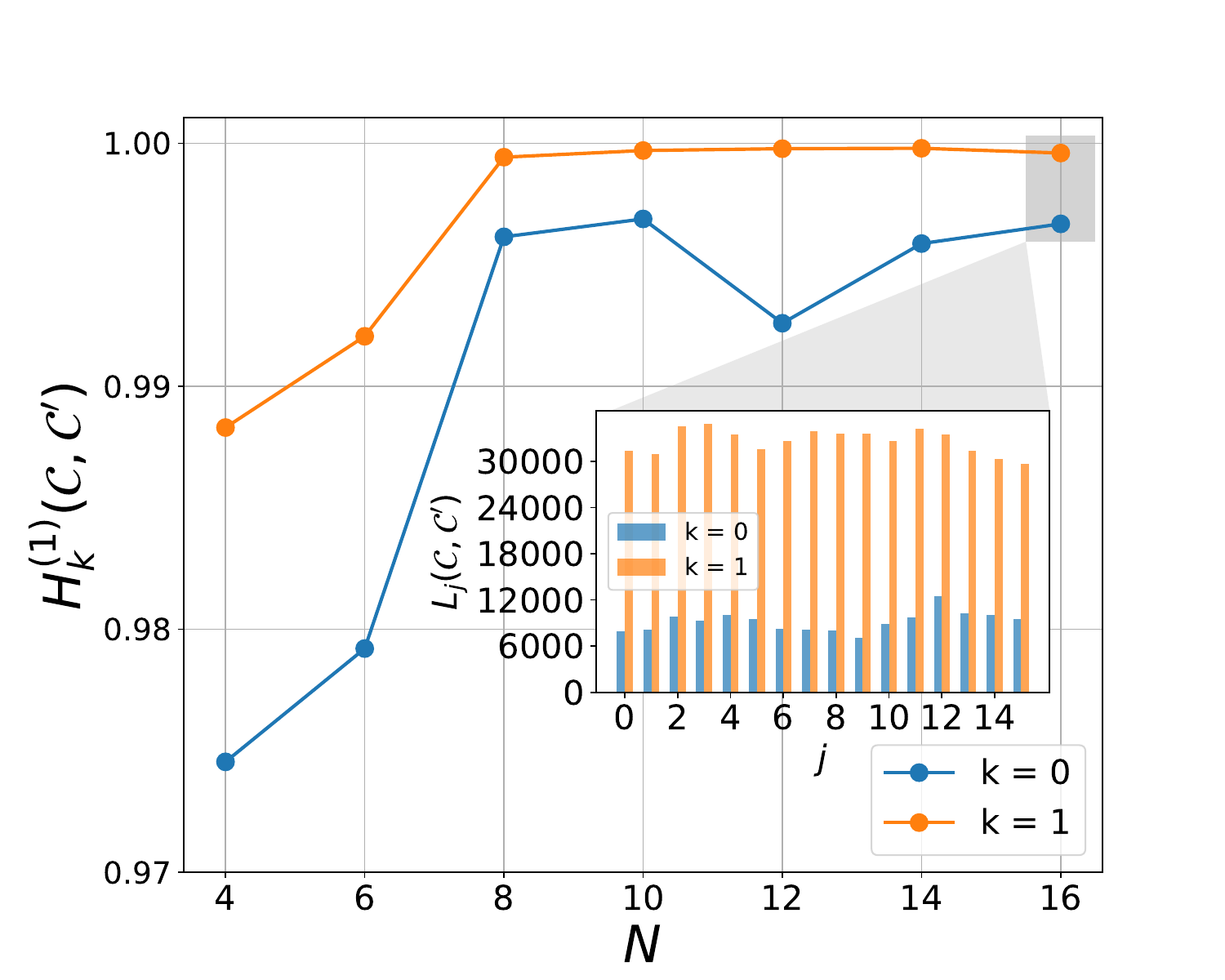}
    \put(-500,190){(a)}
    \put(-250,190){(b)}
    \caption{Diffusion analysis of the quantum hash function. (a) The degree of diffusion indicated by the normalized average Hamming distance $D_k(\mathcal{C}, \mathcal{C}^\prime)$ for various $N$, with error bars representing the standard deviation of the normalized Hamming distance $d_k(\bm{c},\bm{c}')/(N/2)$.
    The inset presents the histograms of the Hamming distance $d_k(\bm{c},\bm{c}^\prime)$ for $N=16$.
    (b) The uniformity of diffusion shown by the normalized Shannon entropy $H_k^{(1)}(\mathcal{C}, \mathcal{C}^\prime)$ for various $N$, measuring the uniformity of bit changes. The inset shows the total number of bit changes $L_j(\mathcal{C}, \mathcal{C}^\prime)$ at each bit position $j$ for $N=16$. In the main panels, the data for different $k$ are slightly shifted horizontally for visibility.
    }
    \label{fig:diffusion}
\end{figure*}
First, we first examine the degree of diffusion using the normalized average Hamming distance $D_k(\mathcal{C}, \mathcal{C}^\prime)$. As depicted in Fig.~\ref{fig:diffusion}(a), the normalized average Hamming distance measures the extent of bit changes across all bits as the output varies for different modes. The error bars represent the standard deviation normalized by $N/2$, indicating the variability in bit changes. Here, $\mathcal{C}^\prime$ represents the set of hash values corresponding to the set of input values $\mathcal{B}^\prime$, where each input value in $\mathcal{B}^\prime$ is obtained by randomly altering one bit of the original input set $\mathcal{B}$. The results show that for $k=0$, there is a high concentration of $D_k$ values near zero, indicating minimal bit flips. 
For $k=1$, the distance approaches the ideal value, reflecting a more consistent diffusion behavior compared to lower values of $k$. Following this, the histogram in the inset of Fig.~\ref{fig:diffusion}(a) presents a snapshot of histogram of Hamming distance $d_k(\bm{c},\bm{c}^\prime)$ for the specific case of $N=16$. The histogram shows that when $k=0$, the distribution has a high concentration near zero, indicating that almost no bit flips occur.
For $k=1$, the histogram exhibits a Gaussian distribution centered around $N/2$ (8 in this case), indicating a balanced degree of diffusion where the number of bit changes clusters around the midpoint.

Second, we analyze the uniformity of diffusion using the normalized Shannon entropy $H_k^{(1)}(\mathcal{C}, \mathcal{C}^\prime)$, as shown in Fig.~\ref{fig:diffusion}(b). The normalized Shannon entropy measures the uniformity of bit changes across all positions. The results indicate that for $k=0$, the entropy values are close to the ideal value, suggesting that the bit changes are uniformly distributed. However, for $k=1$, the entropy values deviate slightly from the ideal value of 1, indicating a less uniform distribution of bit changes in the diffusion case. The chart bar in the inset of Fig.~\ref{fig:diffusion}(b) shows total number of bit changes $L_j(\mathcal{C}, \mathcal{C}^\prime)$ at bit position $j$ for $N=16$. The chart indicates that when $k=0$, the total number of bit flips is low, with some variation across bit positions, suggesting that bit flips are not uniformly distributed. In contrast, for $k=1$, the total number of bit flips is higher, resulting in a more uniform distribution around $2^{N-1}$ of bit changes across different bit positions. This consistent distribution across all bit positions, especially for higher values of $k$, reinforces the entropy analysis and suggests that the quantum hash function's diffusion properties are robust under these conditions.

These two sets of analyses show that our quantum hash function has good diffusion properties.
Also, increasing the parameter $k$ enhances the diffusion property, making the hash function more robust by ensuring that small changes in the input lead to highly unpredictable changes in the hash value. This unpredictability is crucial for maintaining the security and integrity of cryptographic hash functions against potential attacks.

\subsubsection{Collision Resistance}
Collision is a phenomenon in which two distinct inputs produce the same hash output~\cite{Hou2023, Li2018, Yang2016, Zhou2022}, and it becomes a security hole in hash functions for cryptographic applications. Here, we show that collisions occur exponentially rarely, and the average number of attempts to find a collision (i.e., the birthday attack) increases exponentially as $N$ increases.

To evaluate the collision resistance of the quantum hash function, we introduce the normalized collision entropy $H_k^{(2)}(\mathcal{C})$, which is defined as:
\begin{align}\label{eq:beta}
    H_k^{(2)}(\mathcal{C}) &= -\frac{1}{N}\log_2\left(\sum_{\bm{h} \in \{0,1\}^N} P_{\bm{h}}(\mathcal{C})^2 \right), \\
    P_{\bm{h}}(\mathcal{C}) &= \frac{1}{2^N}\sum_{\bm{c} \in \mathcal{C}} \mathbf{I}(\bm{c} = \bm{h}),
\end{align}
where $\mathbf{I}(\bm{c}=\bm{h})$ is the indicator function, which is $1$ if $\bm{c}=\bm{h}$ and $0$ otherwise. The term $P_{\bm{h}}(\mathcal{C})$ represents the proportion of hash values in the set $\mathcal{C}$ that are equal to the bitstring $\bm{h}$. The collision entropy $H_k^{(2)}(\mathcal{C})$ is based on the Rényi entropy of order $2$, which captures the likelihood of collisions in hash values by measuring the uniformity of their distribution. This normalization scales the entropy measure relative to the bit-length $N$ of the hash values, providing a standardized metric for comparing different hash functions. A higher $H_k^{(2)}(\mathcal{C})$ indicates a more uniform distribution of hash values, reflecting stronger resistance to birthday attacks. 
In fact, the number of attempts required for a successful birthday attack, denoted as $N_\mathrm{attack}$, is known to obey the following inequality~\cite{Wiener2005}:
\begin{align}
    0.7 N_\mathrm{bound}(\mathcal{C}) < N_\mathrm{attack} \leq 2 N_\mathrm{bound}(\mathcal{C}),
\end{align}
where
\begin{align}
    N_\mathrm{bound}(\mathcal{C}) = 2^{H_k^{(2)}(\mathcal{C})N/2}
\end{align}
represents an estimated bound on the number of attempts $N_\mathrm{attack}$ required to successfully attack the quantum hash function. By calculating $H_k^{(2)}(\mathcal{C})$,  we can directly estimate $N_\mathrm{attack}$ through $N_\mathrm{bound}(\mathcal{C})$.

\begin{figure}
    \centering
    \includegraphics[width=\columnwidth]{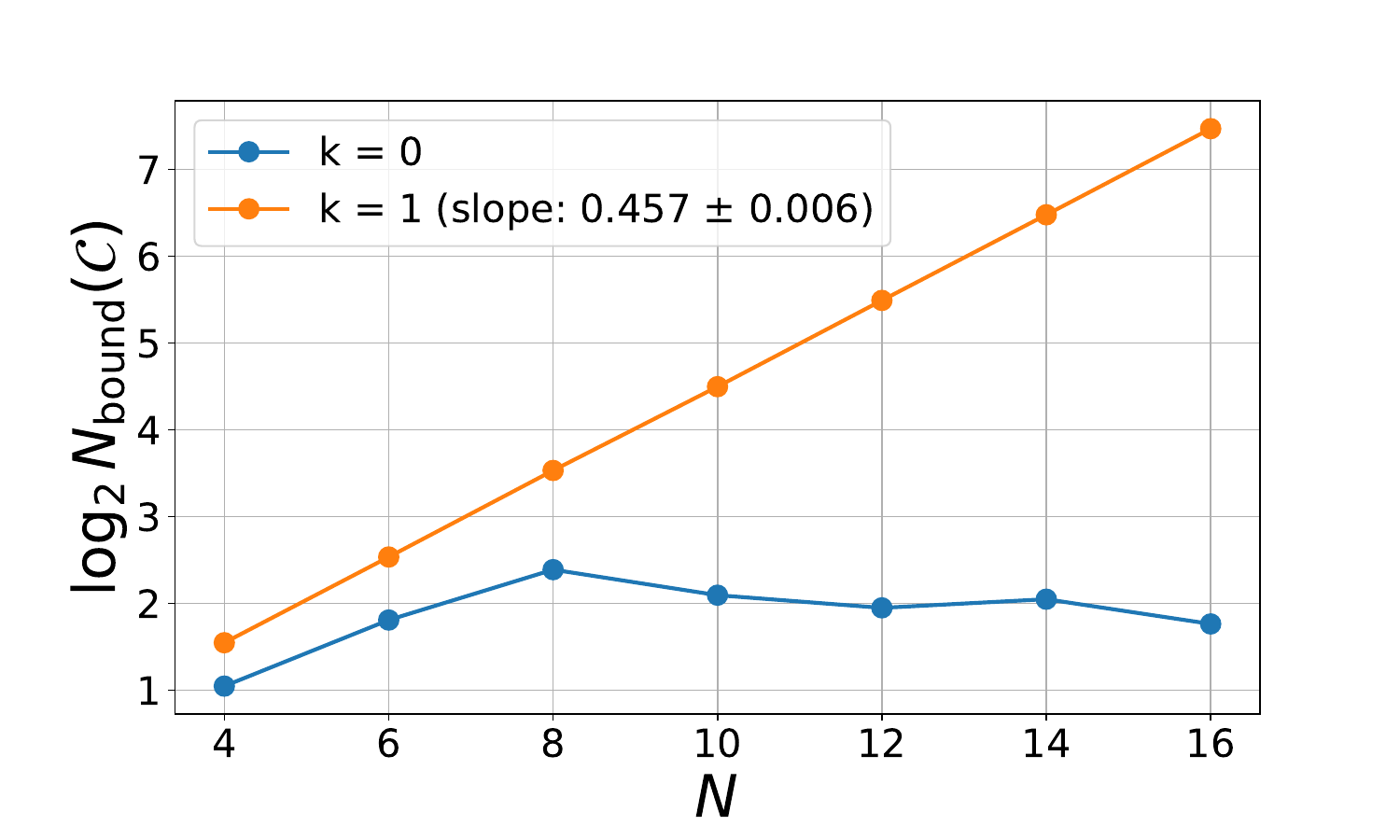}
    \caption{The estimated bound $N_\mathrm{bound}(\mathcal{C})$ versus $N$ for $k=0,1$. The slopes obtained by the least-square fit are $0.457 \pm 0.006$ for $k=1$, close to the ideal value of $0.5$, indicating strong collision resistance.
    }
    \label{fig:collision_entropy}
\end{figure}

Fig.~\ref{fig:collision_entropy} shows $N_\mathrm{bound}(\mathcal{C})$ for different modes as a function of the mode number $N$. For $k=0$, $N_\mathrm{bound}(\mathcal{C})$ decreases as $N$ increases, indicating weaker collision resistance. In contrast, for $k=1$, the slope of $N_\mathrm{bound}(\mathcal{C})$ is approximately $0.457\pm 6\cdot 10^{-3}$ close to the ideal value of $0.5$. This suggests that the quantum hash function exhibits strong collision resistance for these modes. The proximity of these slopes to the ideal value reinforces the robustness of the hash function, indicating that it requires a relatively large number of attempts for a successful collision. To provide further insight into the collision resistance, we also analyzed the distribution of hash values for the case where $N=16$.

Thus, the quantum hash function demonstrates strong collision resistance, especially as the parameter $k$ increases. The distribution of hash values becomes more uniform, and the number of attempts required for a successful birthday attack approaches the theoretical ideal, ensuring the robustness of the hash function against such attacks.

\subsection{Choice of Decimal Parameter $k$}
\label{sec:select_decimal}

We have kept $k$ constant so far and shown that the quantum hash function with $k=1$ works well within the numerically-addressed range of $N\le 16$. Here, we show that $k=1$ seems to keep working well up to a significantly larger $N$ and discuss a guiding principle for choosing $k$ optimally.

A major concern on fixing $k$ as $N$ increases is the possibility that the quantum expectation values $\mu_j$ rapidly decrease. If $\mu_j$ became smaller than $10^{-k}$ for all $j$ and input messages $\bm{b}$, all the hash values would collide as $\bm{c}=\bm{0}$, and the quantum hash function would not work any longer. Thus, it is important to confirm that $\mu_j$ does not decrease rapidly as $N$ increases.

To quantify the magnitude of the three-mode correlations $\mu_j$ over the modes $j$ and the inputs, we introduce their $100x$-upper percentile $\mu_x^\mathrm{th}$, which is defined as
\begin{align}
  \mu_x^\mathrm{th} = \mathrm{arg}\max_\mu \left(F(\mu)\ge x \right)
\end{align}
for a given $0 < x\le 1$, where $F(\mu)$ is a cumulative distribution function which is defined as
\begin{align}
    F(\mu)=\frac{1}{N}\sum_{j=1}^N \frac{1}{2^N}\sum_{\bm{b}\in \mathcal{B}} \Theta(\mu_j(\bm{b}) - \mu)
\end{align}
where $\Theta(y)$ denotes the Heviside step function taking the value $1$ when $y\ge0$ and $0$ when $y<0$. For example, $\mu_1^\mathrm{th}=\min_{j,\bm{b}}\mu_j(\bm{b})$ and $\mu_j(\bm{b})\ge \mu_1^\mathrm{th}$ holds over all modes and inputs. In practice, $\mu_x^\mathrm{th}$ with $x\gtrsim 0.9$ would be used as a reasonable estimate for the magnitude of the three-mode correlations.

The upper percentile $\mu_x^\mathrm{th}$ gives an appropriate decimal parameter $k$ as
\begin{align}\label{eq:kdet}
    k = \left\lceil-\log_{10} \mu_x^\mathrm{th} \right\rceil,
\end{align}
where the ceiling function $\lceil \cdot \rceil$ ensures that $k$ is an integer. 
Roughly speaking, this choice avoids the concern mentioned above with confidence $x$ because \eqref{eq:kdet} implies $\mu_x^\mathrm{th}\ge 10^{-k}$, and hence $\mu_j(\bm{b})\ge 10^{-k}$  for at least a $100x$ percent of $(j,\bm{b})$.

Fig.~\ref{fig:correlations}(a) shows $\mu_x^\mathrm{th}$ for some conservative choices of $100x=$ 99\%, 95\%, and 90\%. These data are consistent with the results that we obtained in the previous subsection in two ways.
First, $\mu_x^\mathrm{th}<1$ for all these choices of $x$, with which \eqref{eq:kdet} tells us $k\ge 1$. We saw that the quantum hash function with $k=0$ did not have good unpredictability properties, whereas $k=1$ worked well.
Second, $\mu_x^\mathrm{th}$ experiences a suppression near $N=6$ as a large finite-size effect. This behavior seems consistent with the worse performance in the confusion and diffusion even for $k=0$ in Figures~\ref{fig:confusion} and \ref{fig:diffusion}.

More importantly, $\mu_x^\mathrm{th}$ does not decrease rapidly as $N$ increases for all choices of $x$ in Fig.~\ref{fig:correlations}(a). Those data for $N\leq 16$ fluctuate and do not show clear downtrends. However, it seems quite reasonable to expect that $\mu_x^\mathrm{th}\gtrsim 10^{-2}$ up to, for instance, $N=256$, which corresponds to the hash size of the SHA-256 standard. This implies that the sampling cost to operate our quantum hash function does not increase very much, as discussed below.

\subsection{Sampling Cost of Quantum Hash Function}
\label{sec:samplingcost}
In Sections~\ref{sec:unpredictability} and \ref{sec:select_decimal}, we assumed an ideal situation in which the quantum expectation values $\mu_j(\bm{b})$ are obtained without statistical errors. When we compute $\mu_j(\bm{b})$ using physical interferometers, however, we measure the output state $\ket{\Psi_\mathrm{out}(\bm{b})}$ in the Fock basis many times, estimating $\mu_j(\bm{b})$ as the sample averages. In this subsection, we estimate the number of samplings (i.e., shots) required, denoted as $N_\mathrm{shot}$.

\begin{figure}
    \includegraphics[height=0.65\columnwidth]{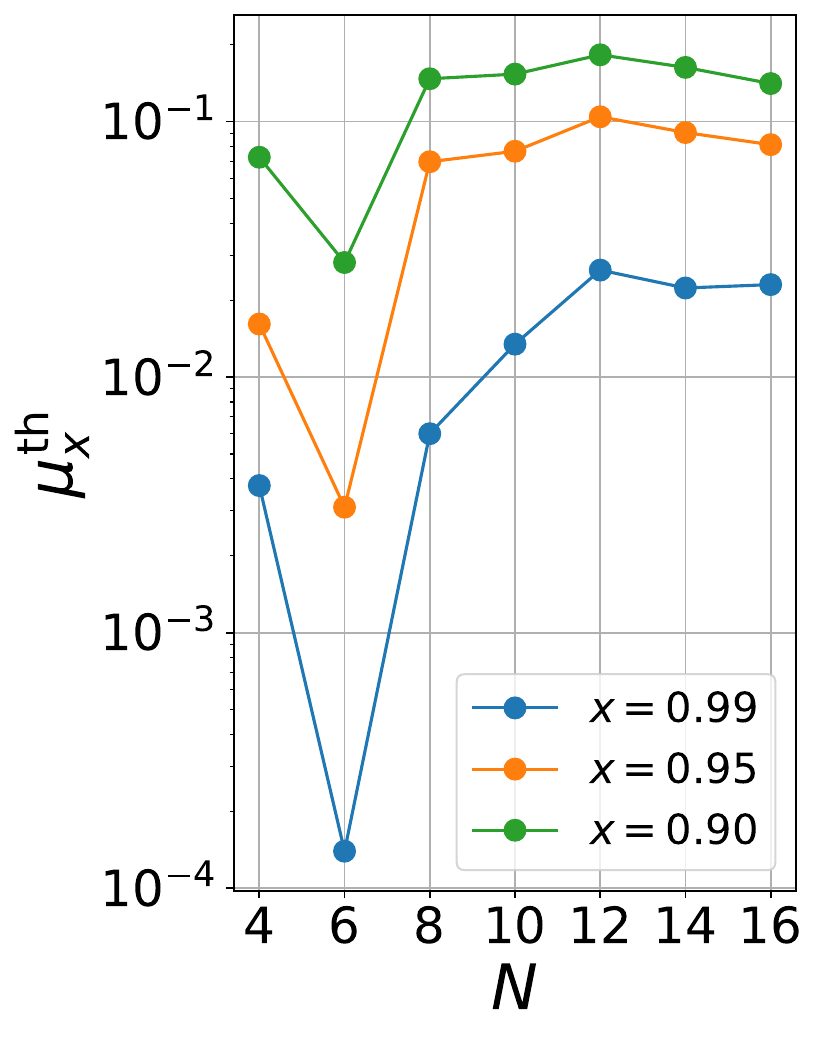}
    \includegraphics[height=0.65\columnwidth]{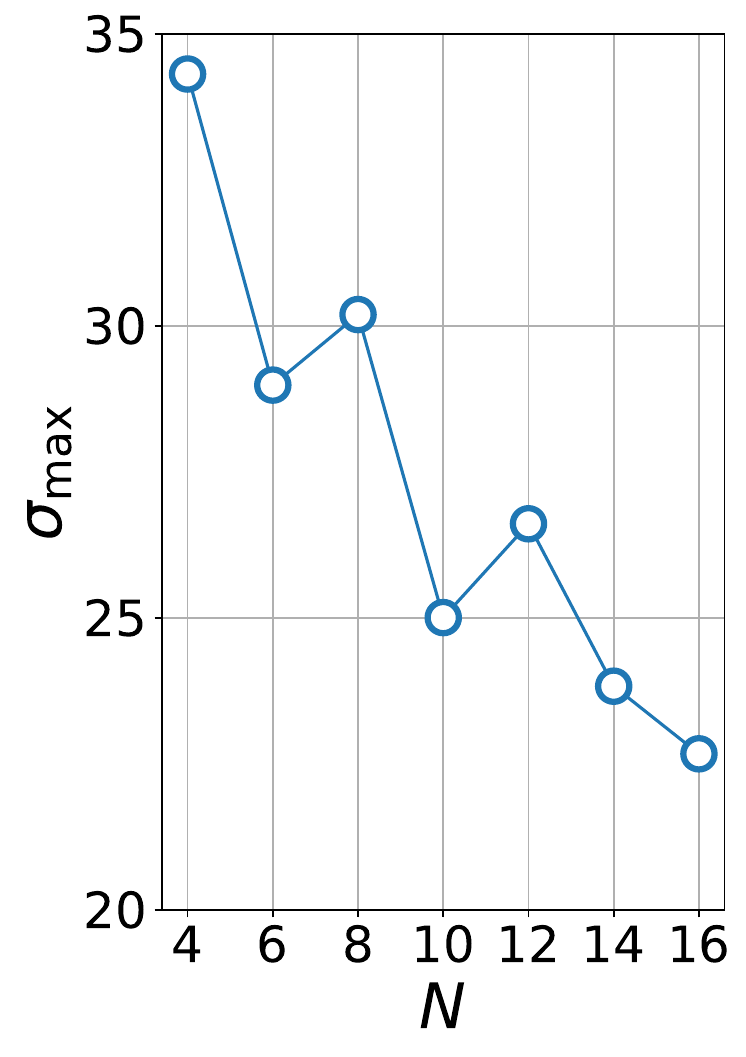}
    \put(-230,160){(a)}
    \put(-105,160){(b)}
    \caption{(a) The upper percentile values of the three-mode correlations $\mu_x^\mathrm{th}$ versus $N$. (b) The maximum intrinsic quantum fluctuation $\sigma_\mathrm{max}$~\eqref{eq:sigma_max} versus $N$.
    }
     \label{fig:correlations}
\end{figure}

The required number of shots, $N_\mathrm{shot}$, depends on the estimation error target $\epsilon$ and the quantum fluctuation of $C_j$. As discussed above, we choose 
\begin{align}
    \epsilon = 10^{-k},
\end{align}
depending on $k$ used in the quantum hash function.
Neglecting systematic errors due to imperfections of devices, we expect that the estimation error of $\mu_j(\bm{b})$ is given by $\sigma_j^\mathrm{est}(\bm{b})=\sigma_j(\bm{b})) /\sqrt{N_\mathrm{shot}}$, where
\begin{align}\label{eq:sigmaj}
    \sigma_j(\bm{b}) = \sqrt{\langle C_{j}^2 \rangle_{\bm{b}} - \langle C_{j} \rangle_{\bm{b}}^2}
\end{align}
denotes the intrinsic quantum fluctuation.
We also take account of the confidence interval, which is controlled by a new parameter $z$ (cf.~the $z$-score). We require $z\sigma_j^\mathrm{est}(\bm{b})\le \epsilon$, ensuring the precision of the estimation with confidence given by $z$. Assuming a normal distribution, $z=1, 2, 3$ correspond to confidence levels of approximately 68\%, 95\%, and 99\%, respectively. Solving $z\sigma_j^\mathrm{est}(\bm{b})\le \epsilon$ for $N_\mathrm{shot}$, we have $N_\mathrm{shot}\ge (z \sigma_j(\bm{b})/\epsilon)^2$ for each $(j,\bm{b})$. Requiring that this inequality holds for all $j$ and $\bm{b}$, we obtain the sufficiently large number of shots as
\begin{align}\label{eq:Nshotmin}
    N_\mathrm{shot}^\mathrm{max} = \left( \frac{z\sigma_\mathrm{max}}{\epsilon}  \right)^2 = z^2 10^{2k}\sigma_\mathrm{max}^2,
\end{align}
where
\begin{align}\label{eq:sigma_max}    
    \sigma_\mathrm{max} =\max_{j=0,\cdots,N-1} \max_{\bm{b}\in \mathcal{B}} \sigma_j(\bm{b}).
\end{align}
This estimation with $\sigma_\mathrm{max}$ is the most conservative, and one can replace $\sigma_\mathrm{max}$ by an appropriate upper percentile value as we did for $\mu_x^\mathrm{th}$.

Equation~\eqref{eq:Nshotmin} enables us to estimate the sampling cost once we know $\sigma_\mathrm{max}$.
Fig.~\ref{fig:correlations}(b) illustrates the $N$-dependence of $\sigma_\mathrm{max}$. Nicely, $\sigma_\mathrm{max}$ gradually decreases as $N$ increases, and $\sigma_\mathrm{max}$ is of the order of 10.
Thus, we obtain $N_\mathrm{shot}^\mathrm{max}\approx 2\times 10^5$ for a reasonable set of $(z,k)=(2,1)$ and $\sigma_\mathrm{max}\approx 23$ at $N=16$, and this number of shots are within the state-of-the-art photonic computers~\cite{Zhong2020}. When we require more confidence (larger $z$) or enhance unpredictability properties (larger $k$), the required number of shots increases according to \eqref{eq:Nshotmin}.

This analysis provides practical insights into the implementation of quantum hash functions, particularly regarding the computational resources and time needed to generate secure hash values.

\section{Application to Blockchain}
Hash functions are fundamental and have many applications in cryptography. Here, we discuss how our quantum hash function applies to blockchains and enhances their security and possibly energy efficiency.
Currently, blockchain systems rely on Proof of Work mechanisms to create new blocks, utilizing hash functions in the process. The predominant hash function employed is the SHA-256 standard, which generates outputs that appear random for any given input. If the output corresponding to a certain input has its first arbitrary number of digits as zeros, the input-output pair is considered valid for the generation of a new block. This is possible because hash functions are structured in such a way that brute-force attacks are the only viable method for searching for an input that yields a desired output. However, it is anticipated that with the advent of quantum computers, Grover's algorithm will provide a quadratic speedup~\cite{Amy2016}, reducing the security level of existing hash functions. Consequently, our quantum-resistant hash function has the potential to become a new paradigm in secure computing.

The GBS approach used in our quantum hash function has already been implemented on a large scale, with claims of quantum advantage being made~\cite{Zhong2020,Zhong2021,Madsen2022,Dellios2023,Deng2023}. While it has been shown that efficient classical simulations are possible when the input suffers from significant decoherence-induced errors~\cite{Rahimi2016,Qi2020}, the quantum advantage is still believed to be achievable if errors are sufficiently minimized. For instance, Madsen et al.~\cite{Madsen2022} demonstrated the use of up to 219 photons to solve a GBS problem that would require 9,000 years on a classical computer but only $36 \text{$\mu$s}$ on a photonic processor~\cite{Zhong2021}. By leveraging quantum advantage, we can constrain the devices capable of generating valid hash functions, thereby increasing the cost of brute-force attacks that require multiple forward calculations. 

Furthermore, it is known that reversible quantum circuits can be constructed for existing classical hash functions like SHA-256~\cite{Amy2016}. Although quantum resource requirements of known implementations~\cite{Amy2016} are beyond the current technology, the existence of such reversible circuits compatible with Grover’s algorithm could benefit from quantum speedup, threatening the hash function security. In contrast, our method involves irreversible measurements and mapping quantum states onto a $2^{256}$-dimensional Hilbert space during the computation process, leading to a significant increase in the dimensionality of the unitary matrix required by Grover's algorithm. This inherent complexity suggests that our hash function possesses quantum resistance. Notably, the backward computation would demand more qubits than the forward computation, indicating that quantum hash functions could achieve virtually perpetual quantum resistance by simply increasing the number of input bits.

\section{Discussions and conclusions}
In this work, we proposed a quantum hash function based on the GBS on a photonic quantum computer. Through extensive simulations, we demonstrated that this quantum hash function exhibits strong properties of confusion, diffusion, and collision resistance, which are essential for cryptographic security.

The degree and uniformity of confusion were shown to approach ideal values as the decimal parameter $k$ increased, indicating enhanced unpredictability. Similarly, the diffusion property, quantified by the impact of input changes on the hash output, also improved with larger $k$ values. Importantly, the estimated number of attempts required for a successful collision attack increased exponentially with the input size $N$, suggesting robust resistance against birthday attacks.

We also analyzed the sampling cost associated with physically implementing the quantum hash function. The required number of measurement shots to achieve a desired precision scaled as $10^{2k}$, almost independently of $N$. This provides valuable insights into the quantum computational resources needed for practical realization.

A key advantage of our quantum hash function lies in its potential application to blockchain technologies. By leveraging the inherent quantum nature of the hash computation, our approach could provide quantum-resistant security to blockchain networks. The high dimensionality of the quantum state space involved in the hashing process poses significant challenges for quantum attacks like Grover's algorithm, indicating a path towards perpetual quantum security with increasing input sizes.

Looking ahead, experimental demonstrations of our quantum hash function on photonic devices would be a crucial step in validating its practical feasibility. Detailed analyses of the robustness against device imperfections and error mitigation techniques will be essential. Integrating our quantum hash function into real-world blockchain testbeds could provide valuable insights into its performance in practical settings.

In conclusion, we have introduced a promising quantum hash function based on Gaussian boson sampling, exhibiting strong cryptographic properties and potential quantum security. Our work lays the foundation for a new paradigm of quantum-resistant hashing with applications in emerging blockchain technologies. With further theoretical and experimental advancements, quantum hash functions could play a pivotal role in safeguarding the integrity and security of future quantum-era information systems.

\section*{Acknowledgment}
The authors thank Sora Akagami and Rumi Hasegawa for their technical support.
M. W. was supported by Grant-in-Aid for JSPS Fellows No.~22KJ1777 and by MEXT KAKENHI Grant No.~24H00957.
T. N. I. was supported by JST PRESTO Grant No. JPMJPR2112 and by JSPS KAKENHI Grant No. JP21K13852.
Also, we used Strawberry Fields~\cite{Killoran2019, Bromley2020}, an open-source software platform for photonic quantum computing, to perform numerical simulations in this research. We would like to thank the developers for making this software available under the Apache License, Version 2.0.

\appendix
\section{Complexity of Interferometer}\label{app:rave}
The transoformation~\eqref{eq:mU} corresponding to the interferometer defines a unitary transformation $U$ among creation operators as
\begin{align}
    \mathcal{U}^\dag a_j^\dag \mathcal{U} = \sum_{j'} U_{jj'}a_{j'}^\dag.
\end{align}
Here, we confirm that $U$ distributes like the Haar measure on the $N\times N$ unitary matrices since it is a necessary condition for the GBS to be computationally hard.

\begin{figure}
    \centering
    \includegraphics[width=\columnwidth]{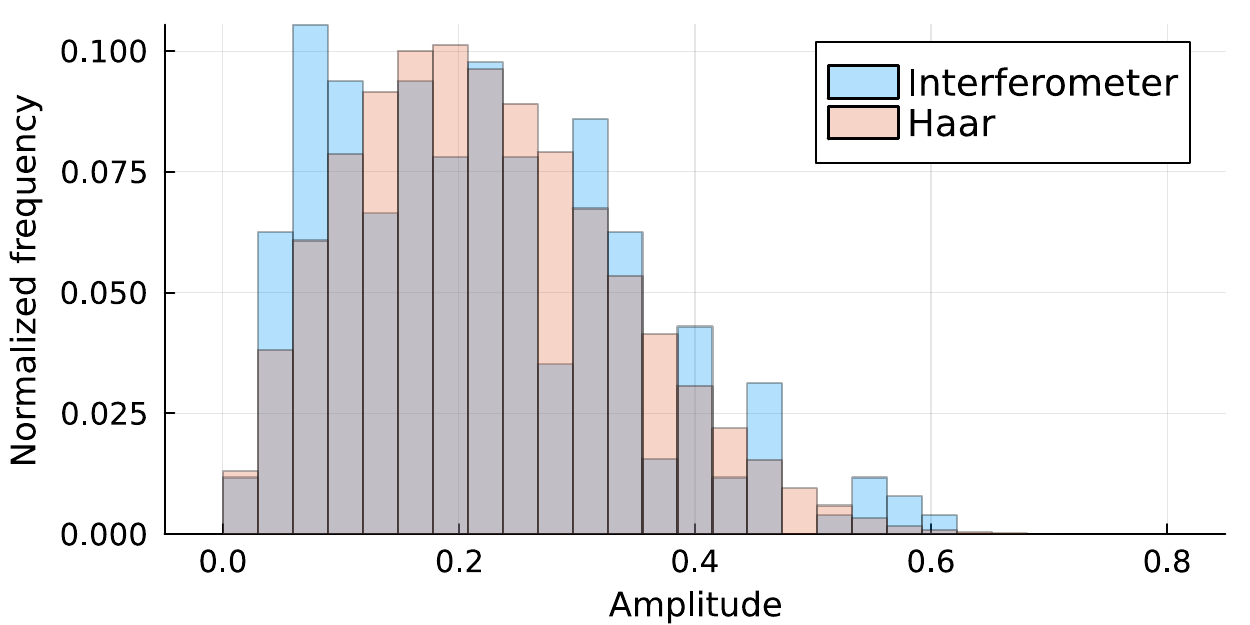}
    \includegraphics[width=\columnwidth]{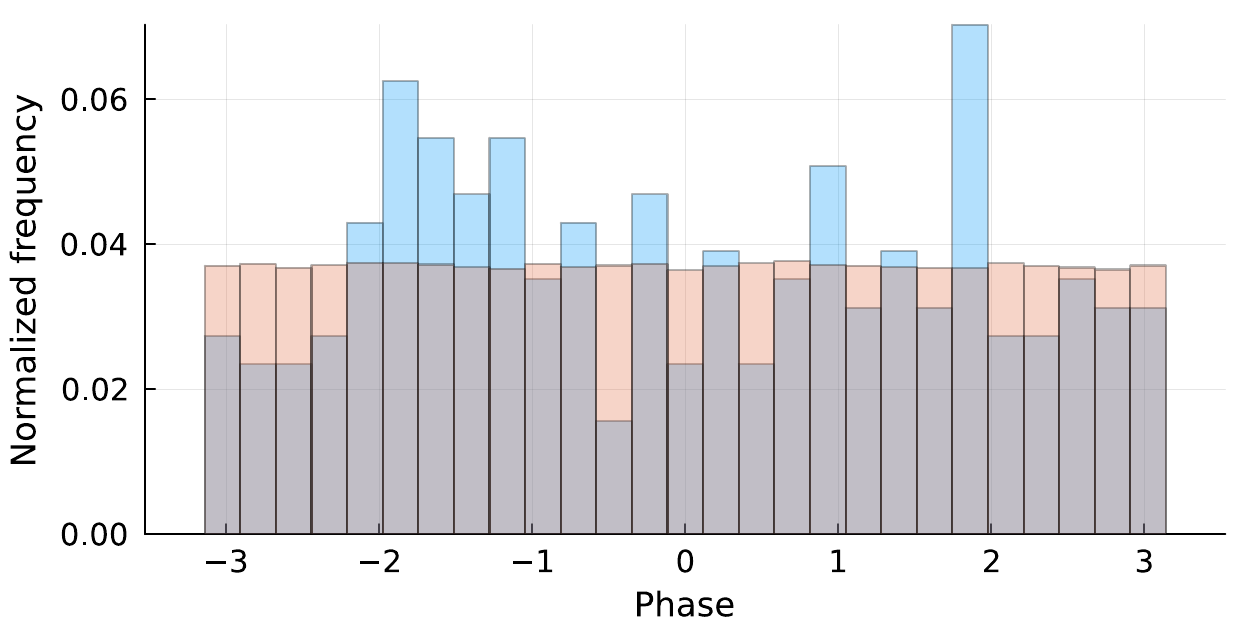}
    \caption{The normalized statistical frequency of the amplitude (top) and phase (bottom) of each element $U_{jj'}$. Blue (pink) bars show the results for our circuit discussed in the main text and the Haar random matrix. The data are taken for depth $d=N=16$ and $10^3$ random realizations for each case.}
    \label{fig:randtest}
\end{figure}

For this purpose, we conduct the Haar test used in the experimental paper~\cite{Zhong2020} and show that our circuit is as close to the Haar measure as the reported one. The test focuses on the distributions of the amplitude and phase of each matrix element $U_{jj'}$. For $N=16$, we averaged over 1000 realizations of the $U$ matrix for each of our circuits and the ideal Haar measure. Fig.~\ref{fig:randtest} compares these distributions normalized so that the sum of the frequencies equals unity. Besides the visual similarities, we confirmed more than $90\%$ overlap between the distributions in terms of the inner product of the normalized frequencies. These results support that our interferometer is so complex that the GBS on it is classically hard.

We also confirmed more than $90\%$ overlap with the Haar measure for the open boundary condition and for the following V-shaped interferometer
\begin{align}
    \mathcal{U}_l &=
        \displaystyle\prod_{j=0}^{N-2}\mathrm{BS}_{j,j+1}(\theta_j^{(l)},\phi_j^{(l)}) \notag\\
        &\qquad \times\displaystyle\prod_{j=N-2}^{0}\mathrm{BS}_{j,j+1}(\bar{\theta}_j^{(l)},\bar{\phi}_j^{(l)}).
\end{align}
Here the newly introduced random variables $\bar{\theta}_j^{(l)}$ and $\bar{\phi}_j^{(l)}$ are all independent and obey the same probability distributions for ${\theta}_j^{(l)}$ and ${\phi}_j^{(l)}$, respectively.
This V shape is easier to realize in actual photonic devices and merits experimental implementations.

\begin{figure*}
    \centering
    \includegraphics[width=2\columnwidth]
    {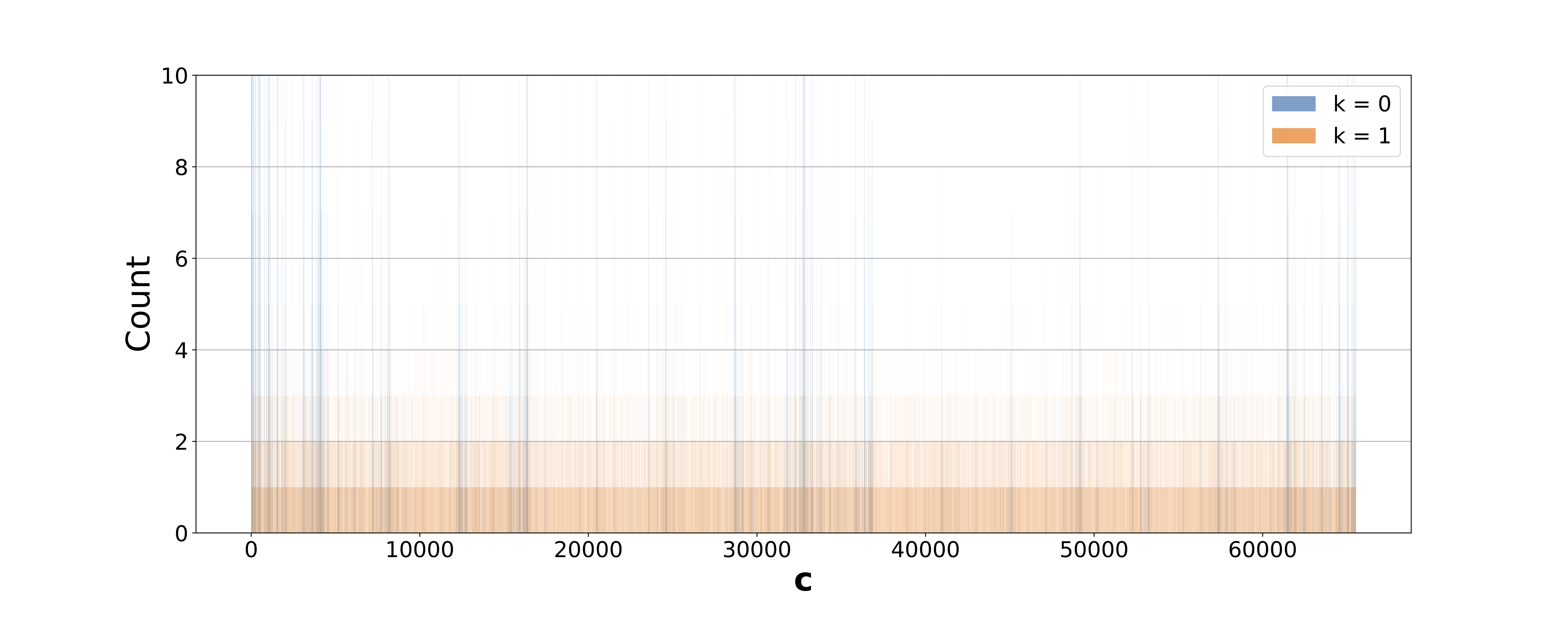}
    \caption{Distribution of hash values for $N=16$. Each bar shows the number of occurrences of each hash output when all the $2^N=2^{16}$ inputs are mapped.}
    \label{fig:collision_distribution}
\end{figure*}

\section{Explicit Hash Value Distributions}
In the main text, we presented the collision entropy for various $N$. Here we provide some raw distributions for $N=16$ as in Fig.~\ref{fig:collision_distribution}.
For $k=0$, there is a higher frequency of identical hash values, suggesting a greater likelihood of collisions. However, for $k=1$, most hash values correspond to unique inputs, demonstrating a higher level of collision resistance.

\end{document}